\documentclass[twocolumn,showpacs,preprintnumbers,superscriptaddress,amsmath,amssymb]{revtex4}

\usepackage[dvipdfmx]{graphicx}

\begin{document}


\title{Spin dynamics in the high-field phases of volborthite}


\author{M. Yoshida}
\altaffiliation[Present address: ]{Max Plank Institute for Solid State Research, Heisenbergstrasse 1, 70569 Stuttgart, Germany}
\affiliation{Institute for Solid State Physics, University of Tokyo, Kashiwa 277-8581, Japan}
\author{K. Nawa}
\affiliation{Institute for Solid State Physics, University of Tokyo, Kashiwa 277-8581, Japan}
\author{H. Ishikawa}
\affiliation{Institute for Solid State Physics, University of Tokyo, Kashiwa 277-8581, Japan}
\author{M. Takigawa}
\affiliation{Institute for Solid State Physics, University of Tokyo, Kashiwa 277-8581, Japan}
\author{M. Jeong}
\affiliation{Laboratoire National des Champs Magn\'{e}tiques Intenses, LNCMI-CNRS (UPR3228), EMFL, UGA, UPS, and INSA, B.P. 166, 38042 Grenoble Cedex 9, France}
\author{S. Kr\"{a}mer}
\affiliation{Laboratoire National des Champs Magn\'{e}tiques Intenses, LNCMI-CNRS (UPR3228), EMFL, UGA, UPS, and INSA, B.P. 166, 38042 Grenoble Cedex 9, France}
\author{M. Horvati\'{c}}
\affiliation{Laboratoire National des Champs Magn\'{e}tiques Intenses, LNCMI-CNRS (UPR3228), EMFL, UGA, UPS, and INSA, B.P. 166, 38042 Grenoble Cedex 9, France}
\author{C. Berthier}
\affiliation{Laboratoire National des Champs Magn\'{e}tiques Intenses, LNCMI-CNRS (UPR3228), EMFL, UGA, UPS, and INSA, B.P. 166, 38042 Grenoble Cedex 9, France}
\author{K. Matsui}
\affiliation{Department of Physics, Sophia University, Tokyo 102-8554, Japan}
\author{T. Goto}
\affiliation{Department of Physics, Sophia University, Tokyo 102-8554, Japan}
\author{S. Kimura}
\affiliation{Institute for Material Research, Tohoku University, Sendai 980-8577, Japan}
\author{T. Sasaki}
\affiliation{Institute for Material Research, Tohoku University, Sendai 980-8577, Japan}
\author{J. Yamaura}
\affiliation{Materials Research Center for Element Strategy, Tokyo Institute of Technology, Yokohama 226-8503, Japan}
\author{H. Yoshida}
\affiliation{Department of Physics, Hokkaido University, Sapporo 060-0810, Japan}
\author{Y. Okamoto}
\affiliation{Institute for Solid State Physics, University of Tokyo, Kashiwa 277-8581, Japan}
\affiliation{Department of Applied Physics, Nagoya University, Nagoya 464-8603, Japan}
\author{Z. Hiroi}
\affiliation{Institute for Solid State Physics, University of Tokyo, Kashiwa 277-8581, Japan}


\date{\today}

\begin{abstract}
We report single-crystal $^{51}$V NMR studies on volborthite Cu$_3$V$_2$O$_7$(OH)$_2 \cdot $2H$_2$O, 
which is regarded as a quasi-two-dimensional frustrated magnet with competing ferromagnetic and antiferromagnetic interactions. 
In the 1/3 magnetization plateau above 28 T, the nuclear spin-lattice relaxation rate 1/$T_1$ indicates an excitation gap 
with a large effective $g$ factor in the range of 4.6-5.9, pointing to magnon bound states. 
Below 26 T where the gap has closed, the NMR spectra indicate small internal fields with a Gaussian-like distribution, 
whereas 1/$T_1$ shows a power-law-like temperature dependence in the paramagnetic state, which resembles a slowing down of spin fluctuations associated with magnetic order. 
We discuss the possibility of an exotic spin state caused by the condensation of magnon bound states below the magnetization plateau.
\end{abstract}


\maketitle



The possibilities of exotic states in quantum spin systems with frustrated interactions have 
attracted strong attention \cite{Frust,Balents}. For example, the ground state of the spin-1/2 
kagome antiferromagnet is believed to show no long-range magnetic order. 
Theories have proposed various ground states such as spin liquids \cite{Yan,Depenbrock,Iqbal} 
or valence-bond-crystal states \cite{Singh}. Other interesting states such as spin nematic 
states \cite{Shannon,Chubukov,Heidrich,Hikihara,Sudan} and magnetization plateaus \cite{Kageyama,Takigawa,Suzuki,Nishimoto,Capponi} 
are also expected in frustrated spin systems. Theories have predicted that a spin nematic state is 
realized near a fully polarized state of a frustrated spin system with competing ferromagnetic (FM) interactions $J_1$ 
and antiferromagnetic (AFM) interactions $J_2$, \cite{Shannon,Chubukov,Heidrich,Hikihara,Sudan}. 
In such systems, a spin nematic state is characterized by the condensation of two-magnon bound states. 
The search for a spin nematic phase has been performed by using high-field NMR near the fully polarized 
state of a quasi-one-dimensional $J_1$-$J_2$ chain magnet LiCuVO$_4$, 
although definitive results are not obtained yet \cite{Buttgen,Orlova}. 

\begin{figure}[t]
\includegraphics[width=0.9\linewidth]{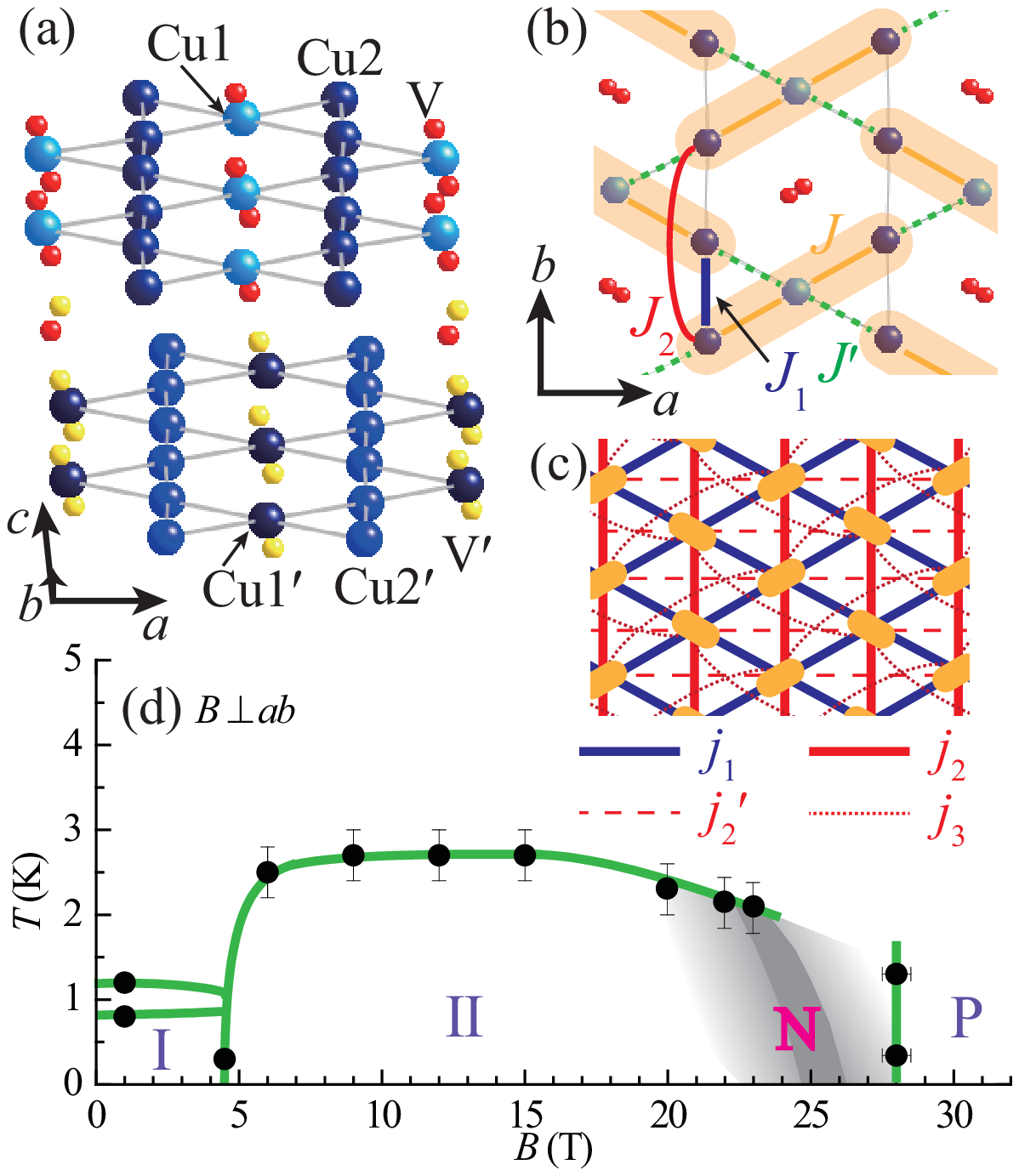}
\caption{(a) Low temperature structure of $P2_1/a$ after the two structural transitions 
at 310 and 155 K \cite{HYoshida2,Ishikawa1,Ishikawa2}. The H and O sites are not shown. 
Although there are two inequivalent layers, the exchange parameters in both layers are expected to 
be nearly the same \cite{Ishikawa2,Janson2} and are described by $J_1$, $J_2$, $J$, and $J^{\prime}$ as shown in (b), 
where the light orange background denotes trimers formed by the dominant coupling $J$. 
(c) Effective model for volborthite suggested in Ref. \cite{Janson2}. 
The orange ellipses represent the trimers carrying pseudospin-1/2. 
The interactions between the effective spins are evaluated 
to be $j_1 = -34.9$, $j_2$ = 36.5, $j_2^{\prime}$ = 6.8, and $j_3$ = 4.6 K \cite{Janson2}. 
(d) Phase diagram of volborthite for the $B \perp  ab$ plane. 
The circles represent the boundaries determined by NMR \cite{Ishikawa2,Suppl}. }
\label{structure}
\end{figure}

Volborthite Cu$_3$V$_2$O$_7$(OH)$_2 \cdot $2H$_2$O is a unique antiferromagnet with frustrated interactions \cite{Hiroi}, 
which contains layers of distorted kagome nets formed by Cu$^{2+}$ ions as shown in Fig.~\ref{structure}(a). 
Early high-field magnetization and NMR measurements in polycrystalline samples revealed three distinct magnetic phases I ($B <$ 4.5 T), 
II (4.5 $< B <$ 26 T), and III (26 T $< B$ ) \cite{HYoshida1,Okamoto,Bert,MYoshida1,MYoshida2,MYoshida3}. 
The magnetic properties were examined on the basis of the distorted kagome model \cite{Wang,Schnyder,Stoudenmire,MYoshida3}, 
while the density functional theory (DFT) study of the $C2/m$ structure proposed that 
frustration should be attributed to the competition between a FM $J_1$ and an AFM $J_2$ between second 
neighbors along the $b$ axis [Fig.~\ref{structure}(b)] \cite{Janson}. 
Recently, single crystals were prepared and they have provided a further opportunity to study the 
unique magnetism of volborthite \cite{HYoshida2,Ishikawa1}. 
In the single crystals, the high-field NMR and magnetization measurements revealed two features remarkably different from 
those previously observed in the polycrystalline samples; one is the 1/3 magnetization plateau (P state) above 28 T 
and the other is the novel phase (phase N) at 23-26 T [Fig.~\ref{structure}(d)] \cite{Ishikawa2}. 

Quite recently, the DFT study of the low temperature structure of $P2_1/a$ 
indicated that the strongest AFM $J$ should lead to an effective model of pseudospin-1/2 
on trimers \cite{Janson2}. The other couplings eventually lead to the realization of a spatially 
anisotropic triangular lattice as shown in Fig.~\ref{structure}(c). Remarkably, this model shows a 
tendency towards the condensation of magnon bound states just below the P state \cite{Janson2}. 

In this paper, we report detailed $^{51}$V NMR studies of the high-field phases in volborthite. 
In the P state, the nuclear spin-lattice relaxation rate 1/$T_1$ indicates an excitation gap 
with a large effective $g$ factor, pointing to magnon bound states. 
In phase N, the NMR spectra indicate small internal fields with a Gaussian-like distribution, 
whereas 1/$T_1$ shows a power-law-like temperature dependence below 2.5 K, which indicates a slowing down of spin fluctuations. 
We discuss the possibility of an exotic spin state caused by the condensation of magnon bound states.  

Two types of crystals A and B were grown by the method described in Refs \cite{Ishikawa1,Ishikawa2}. 
Because volborthite shows a large sample dependence, we discuss this issue in the Supplemental Material 
(see Supplemental Materials A, B, and C \cite{Suppl}). Here, we only show the results of the higher quality crystal A. 
The twinned crystals \cite{Ishikawa1,Ishikawa2} were cut into a single domain for the NMR measurements. 
The data at high magnetic fields above 15 T were mainly obtained by using a 20 MW resistive magnet at LNCMI Grenoble. 
The partial data ($T$-dependences of 1/$T_1$ and spectra at 18-24 T) were obtained by using a hybrid magnet at Tohoku University. 
The NMR spectra were obtained by summing the Fourier transform of the spin-echo signal obtained at equally spaced magnetic fields $B$ 
(or frequencies $\nu$) with a fixed frequency $\nu_0$ (or a fixed field $B_0$). 
They are plotted against the internal field $B_{\mathrm{int}} = \nu_0/\gamma - B$ or $\nu/\gamma - B_0$, where $\gamma $ = 11.1988 MHz/T 
is the gyromagnetic ratio of $^{51}$V. We determined 1/$T_1$ near the spectral center by fitting the spin-echo intensity $M_l(t)$ as a function of the time $t$, 
after a comb of several saturating pulses, to the stretched exponential function 
$M_l(t) = M_{eq} - M_0\mathrm{exp}\{-(t/T_1)^{\beta}\}$, where $M_{eq}$ is the intensity at the thermal equilibrium. 
When the relaxation rate is homogeneous, the value of $\beta$ is close to one.

\begin{figure}[t]
\includegraphics[width=0.9\linewidth]{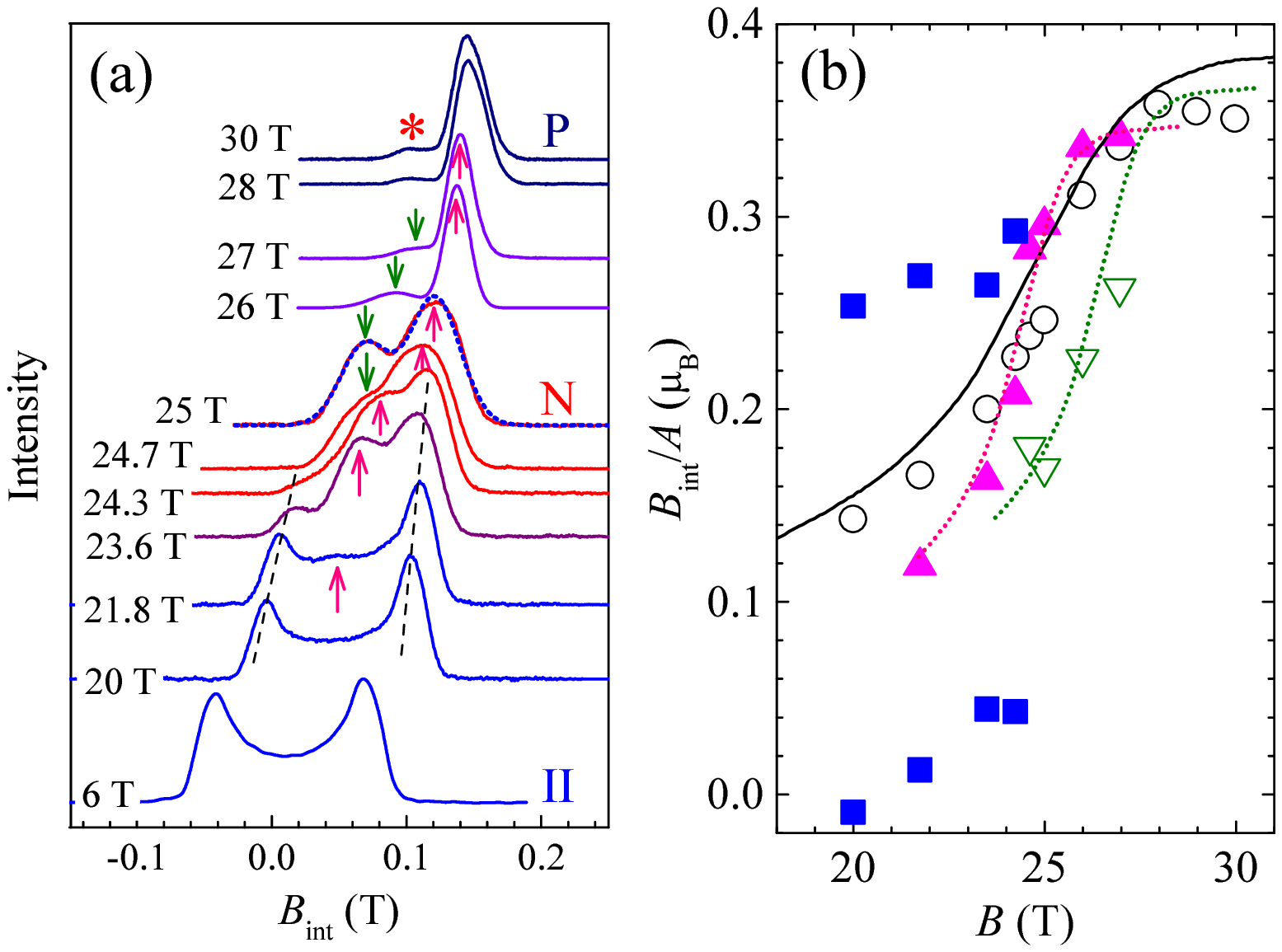}
\caption{(a) Field dependence of the NMR spectra at 0.3-0.4 K in the $B \perp  ab$ plane. 
(b) $B$-dependences of the local spin polarization, $B_{\mathrm{int}}/A$, corresponding to various features on the spectra. 
$A$ = 0.41 T/$\mu_B$ is the hyperfine coupling constant determined in the paramagnetic state. 
The open circles, solid squares, and solid (open) triangles represent the center of gravity, 
the two peaks of the double-horn spectrum indicated by the dashed lines in (a), and the peaks indicated by the up (down) arrows in (a), respectively. 
The solid line represents the magnetization for the $B \perp  ab$ plane \cite{Ishikawa2}. The dotted lines are a guide to the eye.}
\label{spectra}
\end{figure}

We first examine the NMR spectra in order to elucidate the phase diagram. 
Figure~\ref{spectra}(a) shows the $B$-dependence of the NMR spectra at 0.3-0.4 K in the $B \perp  ab$ plane. 
Below 20 T, a double-horn type line shape is observed, which indicates a spin-density-wave (SDW) order \cite{Ishikawa2}. 
Above 22 T, the double-horn structure is deformed and an additional peak grows, indicating the coexistence of phase II and N. 
The spectrum at 25 T can be well fit to two Gaussians, as shown by the dotted line. The two-peak structure seems to be a characteristic 
feature of phase N. Above 26 T, the right peak becomes much narrower, while the left peak remains broad. 
The intensity of the broad peak decreases at 28 T, but it remains visible as indicated by the asterisk. 

The sharp peak observed in the P state indicates a simple spin structure. 
The plateau state is described by the saturation of effective spin-1/2 moments in the coupled trimer model \cite{Janson2}. 
Because $B_{\mathrm{int}}$ at the V sites is unique in this saturation state, 
it is compatible with the observed NMR spectrum, except for the broad peak indicated by the asterisk in Fig.~\ref{spectra}(a), 
which may originate from an imperfection of the crystal. 

We summarize the phase diagram for the $B \perp  ab$ plane in Fig.~\ref{structure}(d). 
The regions of phases I and II are almost the same as those in the polycrystalline sample \cite{Suppl}. 
It is difficult to specify the phase boundaries for N, because it has coexistence regions with phase II and the P state. 
The gray area indicates $B$- and $T$-ranges where a component of phase N was observed. 
The obtained phase diagram is similar to that expected in an ideal two-dimensional model of the spatially anisotropic 
triangular lattice \cite{Starykh}, supporting the validity of the model shown in Fig.~\ref{structure}(c). 
Furthermore, the analyses of the magnon spectra that take into account the longer range $j_2^{\prime}$ 
and $j_3$ couplings indicate a competition between the critical fields related to one- and two-magnon gaps \cite{Janson2}. 
Therefore, it is important to determine which magnetic state is realized below the plateau. 

\begin{figure}[t]
\includegraphics[width=0.8\linewidth]{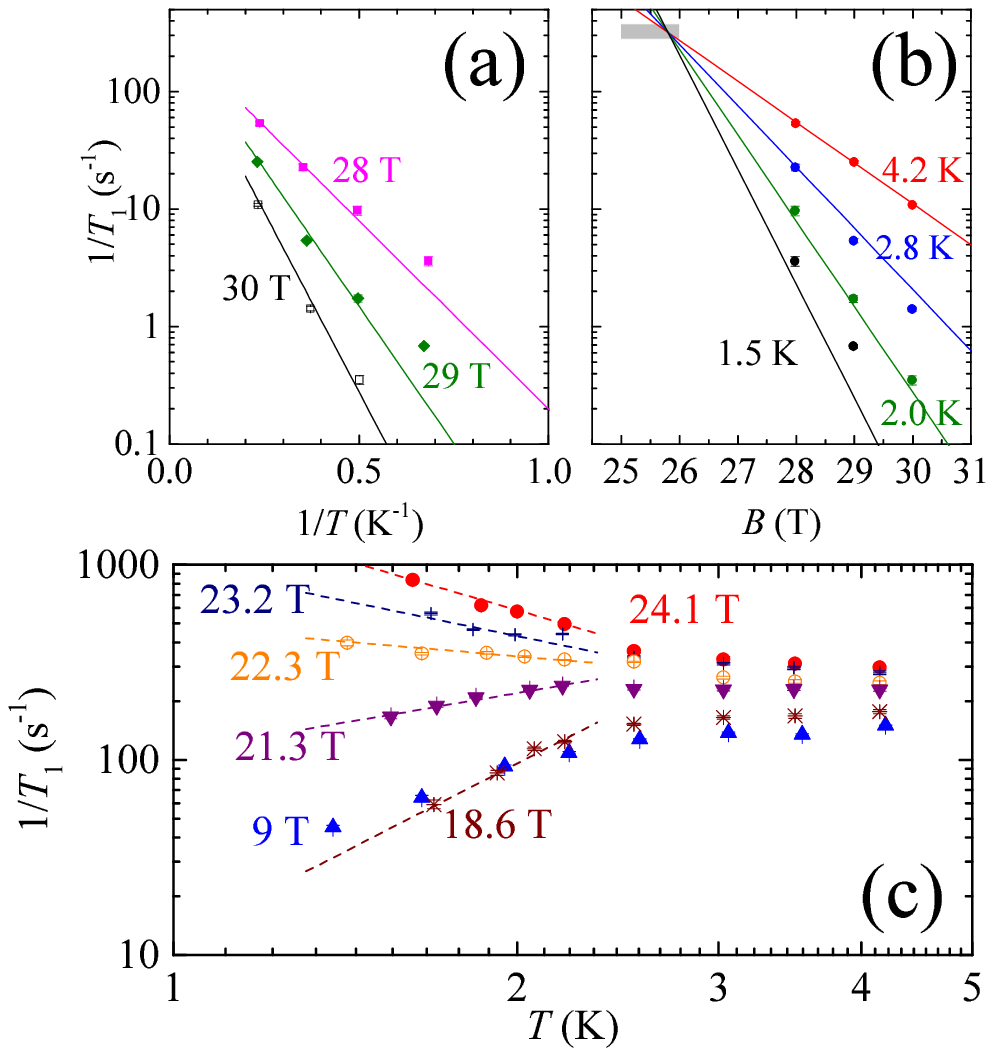}
\caption{
(a) Inverse temperature, (b) magnetic field, and (c) temperature dependences of 1/$T_1$. The data in (a) and (b) are identical.  
1/$T_1$ was measured at the peak position in the plateau region. 
The solid and dashed lines are the fit to $1/T_1 = C\mathrm{exp}[-g\mu_B(B-B_c)/k_BT]$ and $1/T_1 \propto  T^{\alpha}$, 
respectively.
}
\label{dynamics}
\end{figure}

Figure~\ref{dynamics} (a), (b), and (c) shows the 1/$T$-, $B$-, and $T$-dependences of 1/$T_1$, respectively. 
The stretch exponent $\beta$ associated with the determination of 1/$T_1$ is shown in Supplemental Material D \cite{Suppl}. 
In the plateau region, 1/$T_1$ should obey an activation law $1/T_1 = C\mathrm{exp}[-g\mu_B(B-B_c)/k_BT]$, 
where $B_c$ is the critical field. Indeed, 1/$T_1$ above 28 T shows an exponential dependence, as shown in Figs. 3(a) and 3(b). 
It should be noted that the two components are observed in the recovery curve above 28 T (see Supplemental Material E \cite{Suppl}) 
and only 1/$T_1$ data for the slow relaxation component are shown in Figs. 3(a) and 3(b). The origin of the two components is discussed later. 
The $B$-dependence of 1/$T_1$ shown in Fig. 3(b) is directly related to the $g$ factor. 
However, the uncertainty of $B_c$ causes an error in the $g$ factor due to the small number of the data points. 
Therefore, we first investigate the range of the parameters. 
The critical field $B_c$ should be in the range of 25-26 T, because the sharp peak indicating the plateau state 
appears above 26 T as shown in Fig. 2 (a). The constant $C$ is expected to be 330 $\pm $ 50 s$^{-1}$ 
from the extrapolation of the $B$-dependence of 1/$T_1$ at the lower fields. 
When $C$ and $B_c$ are restricted in this area indicated by the gray shade in Fig. 3(b), the $g$ factor is estimated to be 
in the range of 4.6-5.9,  which is two or three times larger than $g$ = 2.0-2.4 in the paramagnetic state \cite{Ohta}. 
In Figs. 3(a) and 3(b), the solid lines show a reasonable fit with $C$ = 320 s$^{-1}$, $B_c$ = 25.8 T, and $g$ = 5.0. 
Below 1.5 K, the data deviate from the fit, probably due to an imperfection of the sample or 
a long-range contribution from the other component. 
This result suggests that two- or three-magnon bound states are the lowest-energy excitations in the P state and 
the condensation of the bound states leads to phase N below 26 T. 
A similar result has been obtained above the saturation field of LiCuVO$_4$ \cite{Buttgen}. 

There is a possibility that 1/$T_1$ is dominated by the three-magnon process rather than the usual Raman process \cite{BEEMAN}. 
If the three-magnon process is dominant in nuclear relaxation, a single-magnon band provides a double value of the gap energy 
due to an additional Bose factor. However, the contribution of the three-magnon process is one or two orders smaller than that 
of the Raman process in the present $T$-range owing to the larger gap. Only if the off-diagonal terms of 
the hyperfine coupling were negligible would a contribution of the three-magnon process be observed \cite{BEEMAN}. 

On the other hand, the spectrum at 25 T can be reproduced by the two Gaussians as shown in Fig.~\ref{spectra}(a). 
If the two-peak structure is associated with AFM internal fields, we have to discard the possibility of a spin-nematic state in phase N. 
Therefore, it is important to examine the field dependence of the spectra. 
As discussed below, our NMR results are well explained by assuming two {\it distinct} components, each assigned to one Gaussian. 
In Fig.\ref{spectra}(b) one can follow various characteristic features on the spectra. 
The center of gravity (open circles) is reasonably consistent with the magnetization (solid line) \cite{Ishikawa2}. 
The signal ascribed to phase N appears at 21.8 T, as indicated by the up arrow in Fig.~\ref{spectra}(a). 
With increasing $B$, the broad peak shifts to higher values of $B_{\mathrm{int}}$, which is plotted 
by the solid triangles in Fig.~\ref{spectra}(b). In addition, another broad peak appears above 24.7 T, 
as indicated by the down arrows in Fig.~\ref{spectra}(a), whose $B$-dependence is shown by the open triangles in Fig.~\ref{spectra}(b). 
From these features, we can set up a two-component scenario where the two broad peaks 
shift independently toward the 1/3 plateau as shown by the dotted lines in Fig.~\ref{spectra}(b). 
At $B_c$ = 26 T, one component reaches the 1/3 plateau, 
while the other component remains at lower $B_{\mathrm{int}}$ in phase N. 

\begin{figure}[t]
\includegraphics[width=0.8\linewidth]{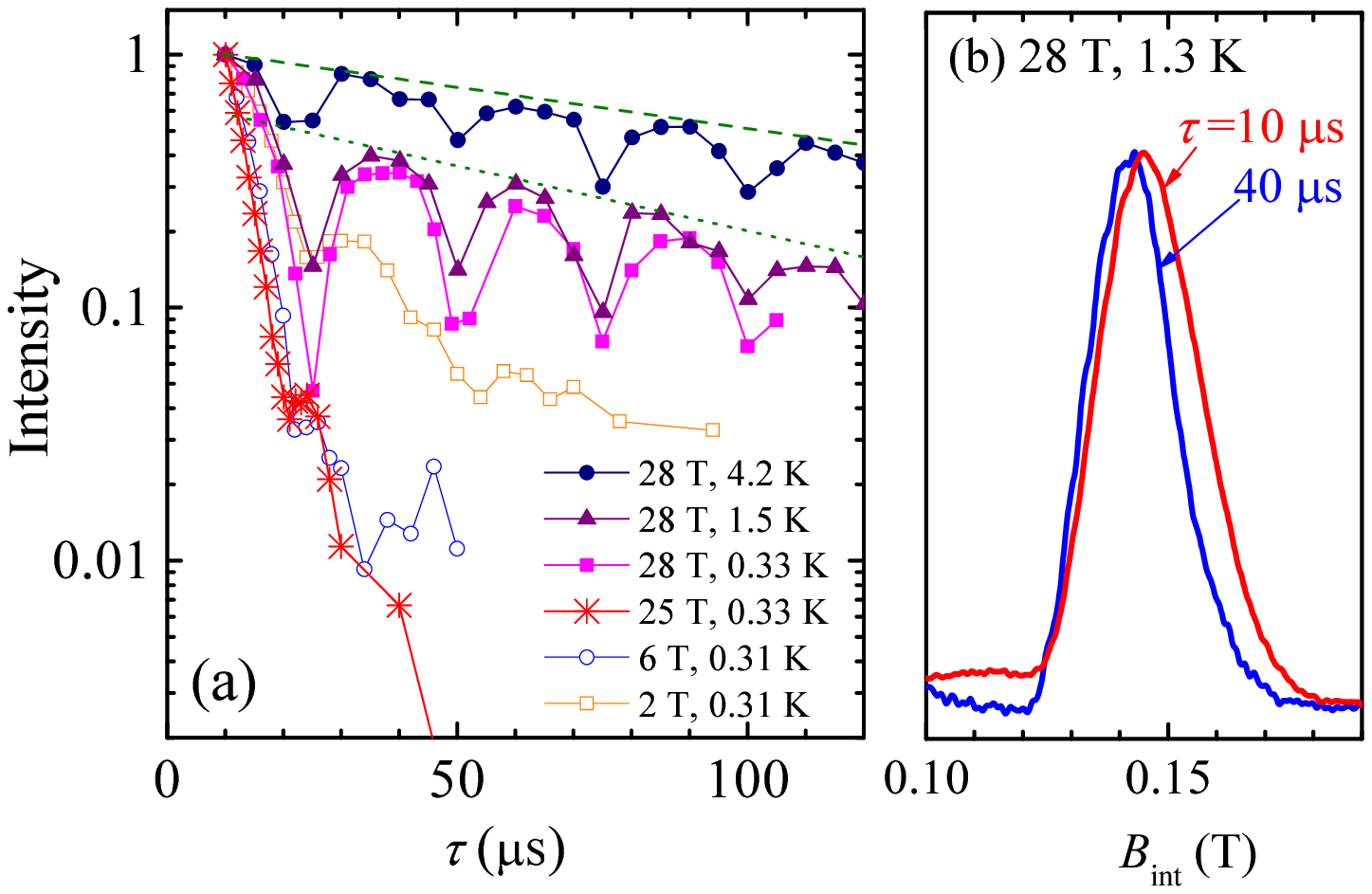}
\caption{(a) Spin-echo decay curves at various magnetic fields in $B \perp  ab$ plane. 
(b) $\tau $ dependences of the NMR spectra at 1.3 K and 28 T in $B \perp  ab$ plane. 
The spectra are normalized by the maximum intensity.
}
\label{echodecay}
\end{figure}

Above 28 T, both components  overlap with each other, resulting in a slightly broader peak compared 
with those at 26 and 27 T.  
The two components can be clearly seen in the spin-echo decay curve. 
Figure~\ref{echodecay}(a) shows the spin-echo decay curves at various magnetic fields and temperatures, 
where $\tau $ is the time between the first and second NMR pulse. 
The echo intensity at 28 T and 4.2 K decreases according to the single exponential function indicated 
by the dashed line. The additional oscillation is caused by quadrupole splitting \cite{Abe}. 
The echo intensities at 1.5 and 0.33 K also decrease according to the single 
exponential function indicated by the dotted line in the region $\tau \geq 30~\mu$s. 
However, the extrapolation of the dotted line to $\tau  = 10~\mu$s shows only 60\% of the actual intensity. 
That is, a component corresponding to 40\% of the intensity disappears until $30~\mu$s. 
Figure~\ref{echodecay}(b) shows the $\tau$ dependence of the NMR spectra at 28 T and 1.3 K. 
The spectral width at $\tau  = 40~\mu$s is smaller than that at $\tau  = 10~\mu$s, 
indicating the two components with different 1/$T_2$ and $B_{\mathrm{int}}$. 
The large difference of 1/$T_2$ between the two components can be attributed to the difference of the critical fields. 
The component with $B_c$ = 26 T has a large excitation gap at 28 T, 
leading to the small 1/$T_2$. The other component must have large 1/$T_2$, because the gap is almost zero at its critical field $B_c^{\prime}$ = 28 T. 

It should be noted that each Gaussian in phase N has a half width at half maximum of 0.02 T, which corresponds to 0.05 $\mu_B$ when scaled by $A$. 
Although an ideal spin-nematic state would not have such inhomogeneous moments, an imperfection of a crystal might induce small moments in real compounds. 
Alternatively, if $n$-magnon bound states ($n \geq  3$) are condensed, a novel SDW state may produce such small internal fields, 
and the rapid increase of $B_{\mathrm{int}}/A$ indicated by the dotted lines in Fig.~\ref{spectra}(b) may indeed 
imply magnon bound states with large $n$ \cite{Sudan,Balents2}. 

Although the origin of the two-component behavior has not been specified yet, 
a possible origin is the two inequivalent layers [Fig.~\ref{structure}(a)]. 
The difference between $B_c$ and $B_c^{\prime}$ can be explained by the small difference in the exchange couplings for the two layers \cite{Janson2}. 
The sample dependence of the magnetization curve may also be explained by the two layer scenario (see Supplemental Material F \cite{Suppl}). 
In this case, the two Gaussians in phase N should have the same intensity. 
The fitting shown by the dotted line in Fig.~\ref{spectra}(a) provides the ratio 1 : 1.6, which may seem to be incompatible with the two layer scenario. 
However, the determination of the actual fraction requires removing the effects of 1/$T_2$, because the values of 1/$T_2$ 
drastically change from phase N to the P state, as shown in Fig.~\ref{echodecay}(a). 

The above analyses indicate that phase N is characterized by the condensation of magnon bound states and a Gaussian-like distribution 
of small internal fields, which seem to support the realization of a spin nematic state or unusual SDW order. 
Nevertheless, at 18-24 T, 1/$T_1$ shows a power-law-like $T$-dependence $1/T_1 \propto  T^{\alpha}$ below 2.5 K 
as shown by the dashed lines in Fig.~\ref{dynamics}(c). Above 22 T, $\alpha$ becomes negative, which seems to indicate a slowing 
down of spin fluctuations towards phase N. Near a spin-nematic transition temperature, an anomaly in 1/$T_1$ is predicted to appear \cite{Smerald}. 
However, this is due to longitudinal fluctuations, which should have only a small contribution. 
The large enhancement of 1/$T_1$ near phase N seems to be incompatible with the condensation of magnon bound states. 
One possible explanation is that unusual slow fluctuations cause this behavior. 
Previous studies have indicated that the spin dynamics in volborthite involves much slower fluctuations than 
the time scale of the NMR frequency ($\sim$100 MHz) \cite{MYoshida1}. 
The spin-echo decay rate 1/$T_2$ is sensitive to such slow fluctuations. 
The decay curves in phases I, II, and N shown in Fig.~\ref{echodecay}(a) provide anomalously large values of 
$1/T_2 \sim$ 0.5, 0.9, and 1.0 $\times 10^5$ s$^{-1}$ at 2, 6, and 25 T, respectively, while the dashed line at 28 T and 4.2 K provides $1/T_2 \approx  4000$ s$^{-1}$, 
which could be explained by contributions from dissimilar nuclear spins \cite{Recchia}. 
These results indicate that unusually slow fluctuations exist in these phases of a high-quality crystal. 
Such slow fluctuations can also contribute strongly to 1/$T_1$ depending on their time scales, 
although the origin of these fluctuations is unclear at present. 
Another possibility is that an exotic phase caused by the condensation of magnon bound states is limited in a small field 
region just below the plateau as in the case of LiCuVO$_4$ \cite{Buttgen,Orlova}. 

In summary, the 1/$T_1$ results in the P state indicate an excitation gap with a large effective $g$ factor, pointing to multi-magnon bound states. 
In phase N, the NMR spectra indicate small internal fields with a Gaussian-like distribution. 
These results support the realization of a spin nematic state or unusual SDW order. 
Nevertheless, at 18-24 T, 1/$T_1$ shows a power-law-like temperature dependence, which indicates a slowing down of spin fluctuations towards phase N.  
We suggested possible scenarios for the magnetic state below the magnetization plateau. 

We acknowledge useful discussions with O. Janson, S. Furukawa, T. Momoi, and J. Richter. 
This work was partly supported by a Grant-in-Aid for JSPS KAKENHI (No. 26800176 and No. 25287083) 
and by EuroMagNET II network under the European Commission Contract No. FP7-INFRASTRUCTURES-228043. 
Part of this work was performed at High Field Laboratory for Superconducting Materials, 
Institute for Materials Research, Tohoku University under the Projects No. 14H0011 and No. 15H0015.

\section*{SUPPLEMENTAL MATERIALS}

\subsection{Sample dependence of magnetization}

\begin{figure}[b]
\includegraphics[width=0.8\linewidth]{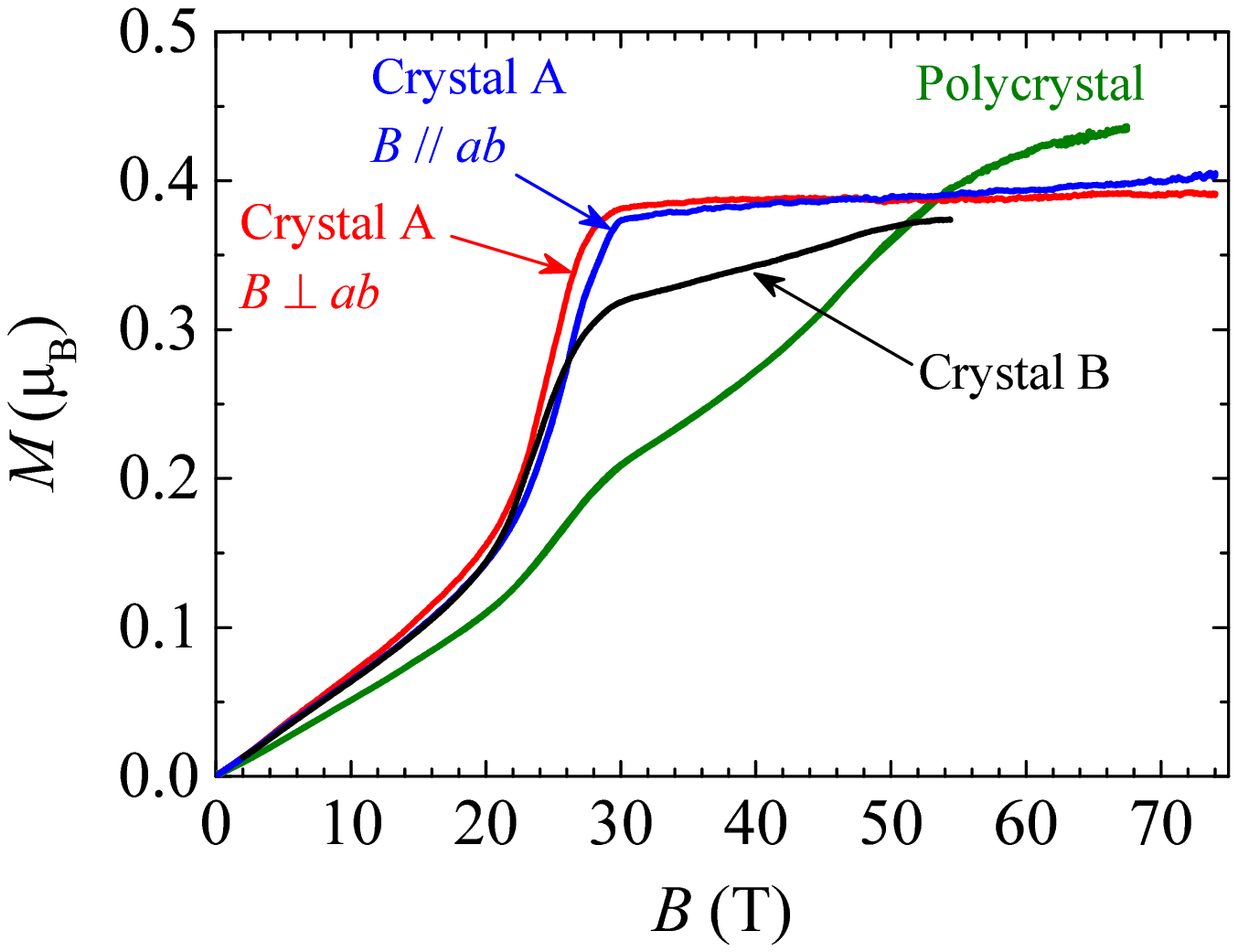}
\caption{Magnetization curves of volborthite for the three samples at 1.4 K. For crystal A, the magnetic field was applied 
perpendicular or parallel to the $ab$ plane \cite{Ishikawa2}. The magnetization for crystal B was measured on a bunch of 
randomly oriented crystals \cite{Unpublished}. The data of the polycrystalline sample is taken from Fef. \cite{Okamoto}.}
\label{magnetization}
\end{figure}

Figure~\ref{magnetization} shows the magnetization ($M$) curves for crystal A \cite{Ishikawa2} and 
the polycrystalline sample \cite{Okamoto} at 1.4 K. The curve for crystal A shows the wide 1/3 magnetization plateau, 
while the curve for the polycrystalline sample shows the step-like increases at 25 and 46 T. 
The step at 25 T in the polycrystalline sample seems to correspond to the rapid increase toward the 1/3 magnetization 
plateau near 25 T in crystal A. The NMR spectrum contains two components at 30 T 
in the polycrystalline sample \cite{Ishikawa2,MYoshida3}. 
One component has larger $B_{\mathrm{int}}$ corresponding to the 1/3 plateau and smaller 1/$T_2$ (pure component). 
The other component has smaller $B_{\mathrm{int}}$ and larger 1/$T_2$ (disorder component). The latter one has a broad Gaussian-like shape, 
suggesting an inhomogeneous distribution of the internal field due to imperfection of the sample \cite{Ishikawa2,MYoshida3}. 
Because the disorder component is nearly absent in crystal A, the crystal must be of higher quality \cite{Ishikawa2}. 

To further investigate the sample dependence, magnetization measurements on a bunch of crystal B 
with random orientations were performed \cite{Unpublished}. 
Here, crystals A and B with the different growth times of 1 month and 2 weeks have the different typical 
sizes of $1 \times  2 \times  0.15$ and $0.5 \times  1 \times  0.05$ mm$^3$, respectively. 
As shown in Fig.~\ref{magnetization}, 
the curve of crystal B shows an intermediate behavior between curves of crystal A and the polycrystalline sample, 
indicating that crystal B contains the disorder component although the amount is smaller than that in the polycrystalline sample. 
This result indicates that the disorder component increases with decreasing the crystal size. 
Imperfection of the crystal would be removed in large crystals (crystal A) due to the long growth time. 

As shown in Fig.~\ref{magnetization}, $M$ for crystal B approaches $M$ for crystal A near 50 T, 
implying that the disorder component also reaches the 1/3 plateau above 50 T. 
The step at 46 T observed in the polycrystalline sample might be attributed to the increase of $M$ of the disorder component 
toward the 1/3 plateau. However, the reason why $M$ for the polycrystalline sample exceeds 
the 1/3 of the saturated magnetization above 52 T is unclear at present. 

Because the magnetization measurements on crystal B were performed by using many small crystals, 
we cannot distinguish two cases; one case is that individual crystals show the same magnetization curve observed 
in the bunch of crystal B and the other case is that there are two types of crystals, that is, 
crystal A and polycrystalline types, in the bunch of crystal B. The NMR results on a single crystal of B below 15 T 
are slightly different from the results of crystal A as shown in the section B and C. 

\subsection{Phase boundaries in the magnetic field perpendicular to the $ab$ plane}

The NMR measurements on the polycrystalline sample revealed the three distinct magnetic phases I ($B <$ 4.5 T), 
II (4.5 $< B <$ 26 T), and III (26 T $< B$ ) \cite{HYoshida1,MYoshida3}. The high-field NMR and magnetization measurements 
on crystal A indicated two phases remarkably different from those in the polycrystalline sample; one is the plateau (P) 
state above 28 T and the other is phase N at 23-26 T \cite{Ishikawa2}. P state and phase N are discussed in the main text. 
In this section, we investigate the phase boundaries in the lower field region, where the difference between 
crystal A and the polycrystalline sample is not significant. 

\begin{figure}[t]
\includegraphics[width=0.7\linewidth]{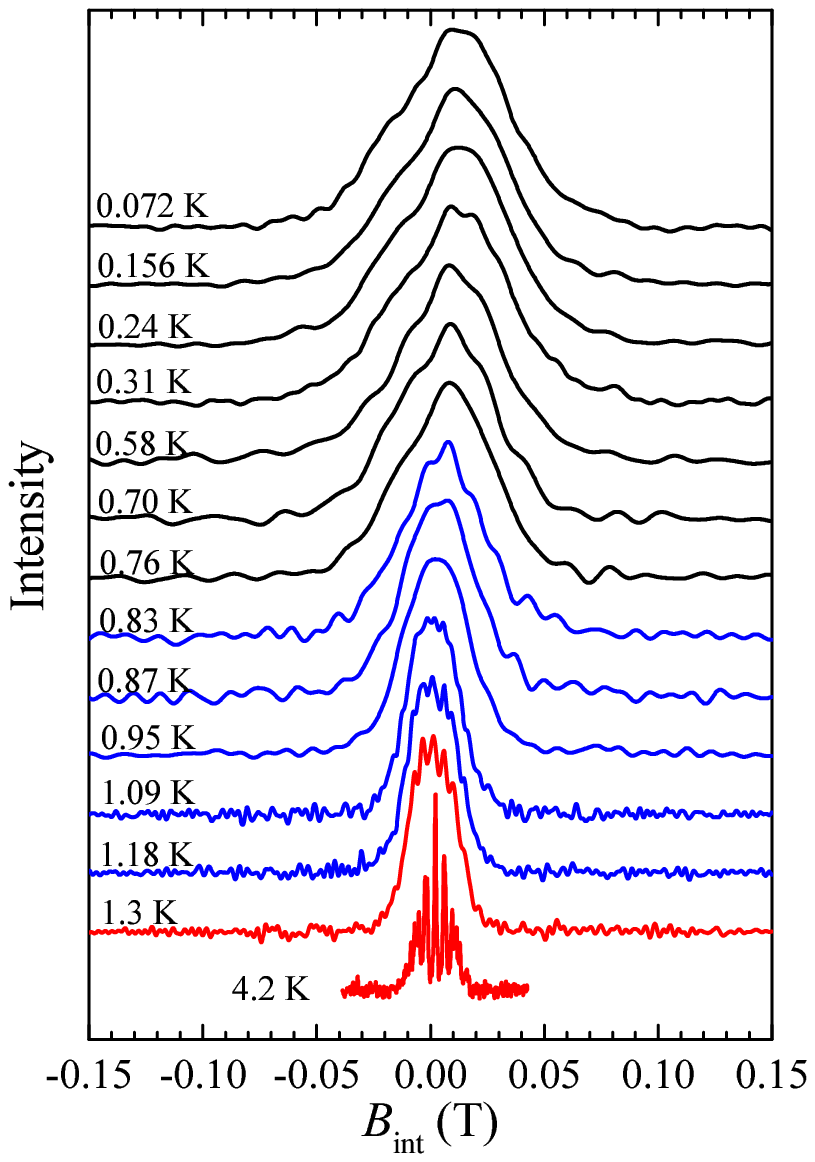}
\caption{Temperature dependence of the NMR spectra for crystal A at 1 T in $B \perp  ab$ plane.}
\label{sp1T}
\end{figure}

Figure~\ref{sp1T} shows the temperature dependence of the NMR spectra for crystal A at 1 T in $B \perp  ab$ plane. 
At 4.2 K, in the paramagnetic region, sharp peaks due to the quadrupole splitting are observed. 
The line width increases with decreasing $T$ and the distribution of $B_{\mathrm{int}}$ masks the quadrupole splitting below 1.2 K. 
The spectra become independent of $T$ below 0.8 K. These results indicate an antiferromagnetic transition near 1 K, 
which is consistent with the results of the polycrystalline sample \cite{MYoshida1}. 
The center of gravity shifts to higher values of $B_{\mathrm{int}}$ below 1 K. 
This shift can be explained by internal fields perpendicular to the external magnetic field. 

\begin{figure}[t]
\includegraphics[width=0.8\linewidth]{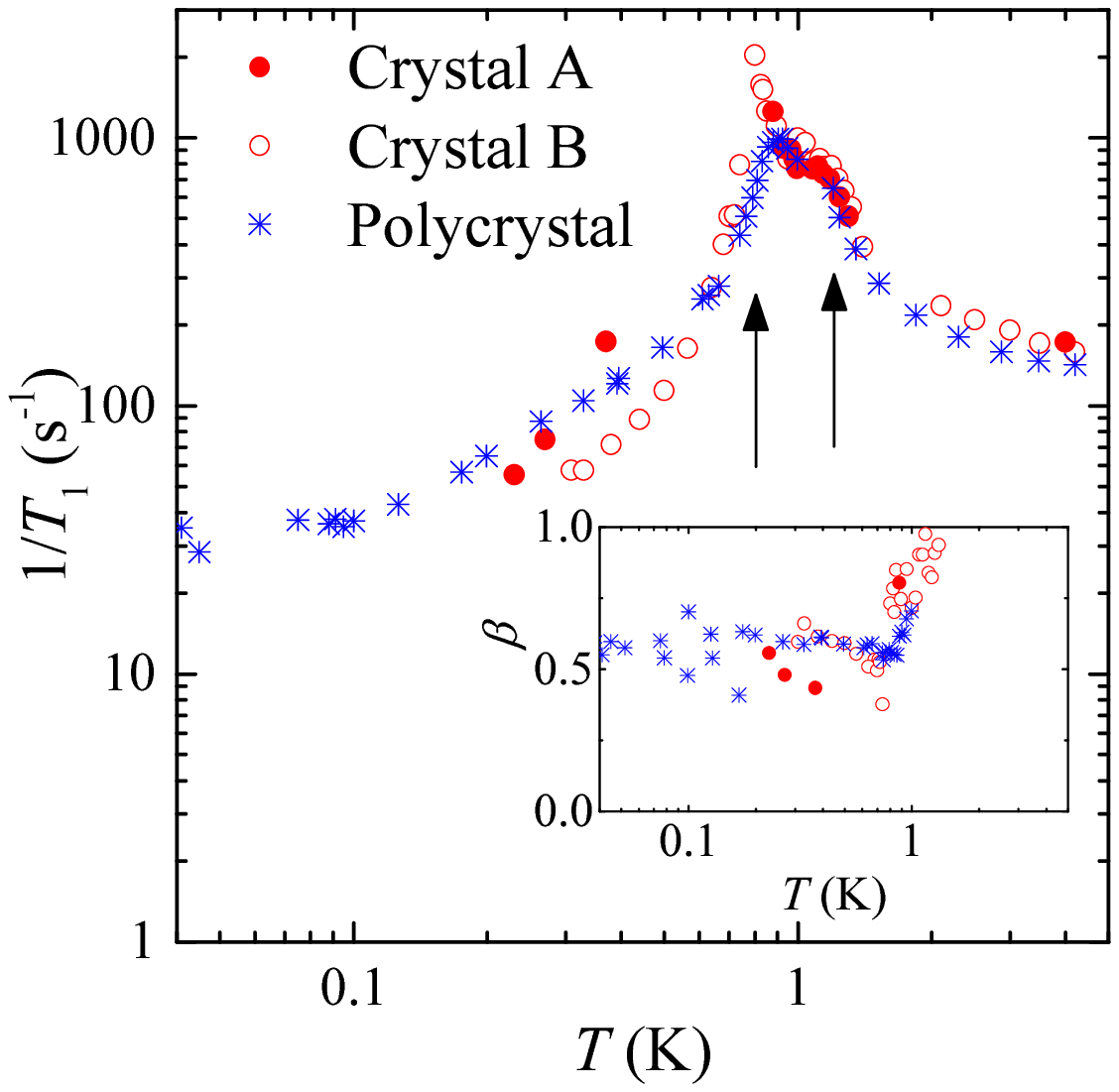}
\caption{Temperature dependences of 1/$T_1$ for the three samples at 1 T. 
For crystal A and B, the magnetic field was applied perpendicular to the $ab$ plane. 
The data of the polycrystalline sample is taken from Ref. \cite{MYoshida1}. 
The inset shows the temperature dependences of the stretch exponent $\beta$ 
We used the single exponential function to determine 1/$T_1$ above 1.3 K.}
\label{T1at1T}
\end{figure}

The heat capacity measurements on single crystals indicate two peaks at 0.8 and 1.2 K \cite{HYoshida2}. 
Although the two-step transition is not clear in the temperature dependence of the NMR spectra, 
two anomalies are observed in the temperature dependence of 1/$T_1$. Figure~\ref{T1at1T} shows 
the temperature dependences of 1/$T_1$ for the three samples at 1 T. The inset shows the temperature dependences 
of the stretch exponent $\beta $, which provides a measure of the inhomogeneous distribution of 1/$T_1$. 
Above 1.2 K, the recovery curve can be fit to the single exponential function. 
The temperature dependences of 1/$T_1$ for the three samples are almost same as shown in Fig.~\ref{T1at1T}. 
The arrows indicate 0.8 and 1.2 K, where the two peaks are observed in the heat capacity. 
1/$T_1$ for crystal B clearly shows a sharp peak at 0.8 K and a small shoulder at 1.2 K. 
The critical divergence at 0.8 K is consistent with the existence of the transverse ordered moments 
indicated by the shift of the center of gravity. 
The divergence behavior is in marked contrast to the behavior at 9 T shown in Fig. 3 of the main text. 

Because the recovery curve for crystal A showed unusual oscillating behavior in the temperature range of 0.5-0.8 K, 
the value of 1/$T_1$ could not be determined in this range. At 0.5-0.8 K, 1/$T_1$ and 1/$T_2$ are expected to be very large, 
so that the nuclear magnetization was not sufficiently saturated by the comb pulse. 
In addition, heating effect should be significant due to short delay times for such large 1/$T_1$. 
The insufficiency of the saturating comb pulse and heating effect might result in the unusual behavior of the recovery curve. 

\begin{figure}[t]
\includegraphics[width=0.7\linewidth]{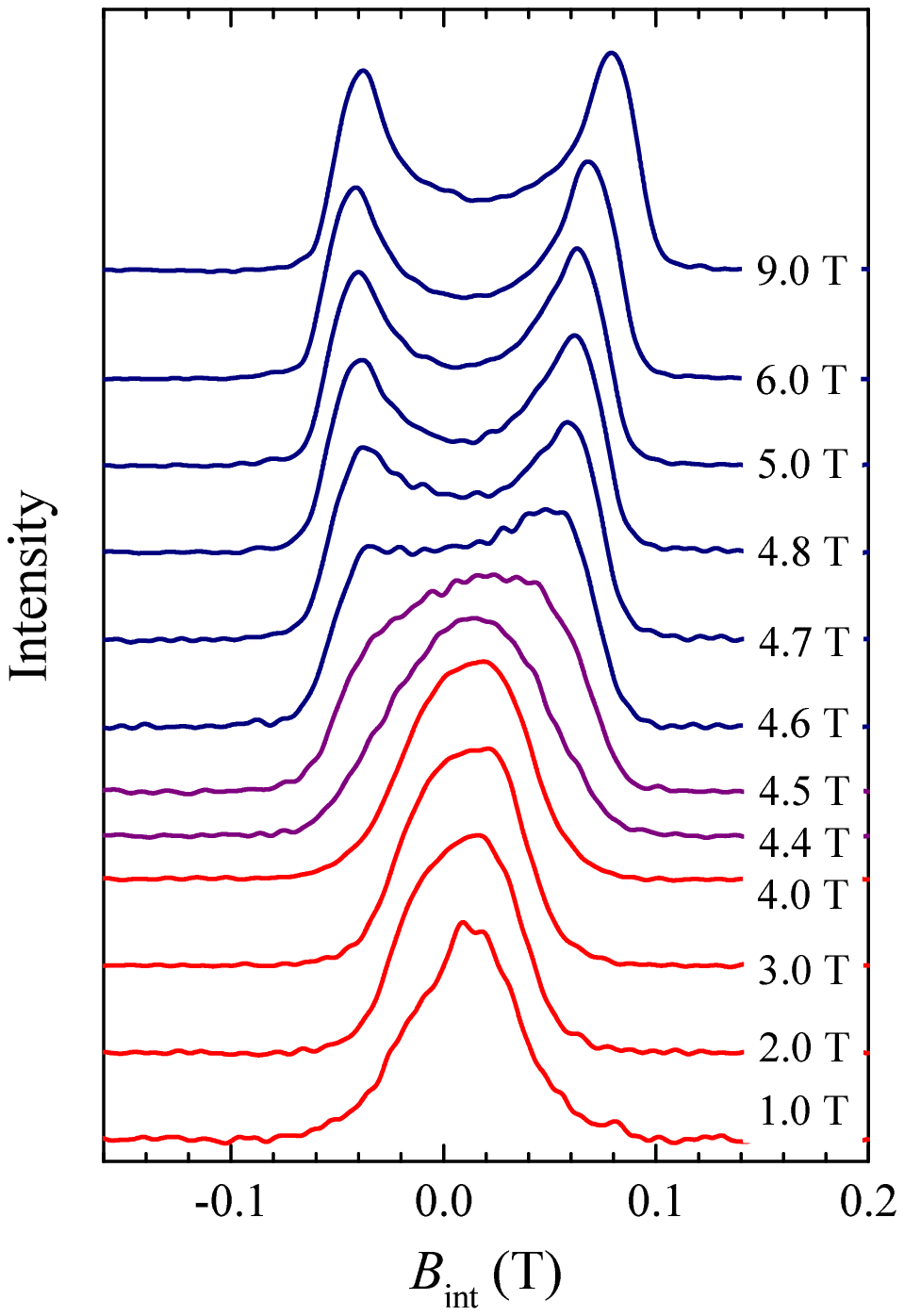}
\caption{Field dependence of the NMR spectra for crystal A at 0.3 K in $B \perp  ab$ plane.}
\label{spFD}
\end{figure}

Figure~\ref{spFD} shows the field dependence of the NMR spectra for crystal A at 0.3 K in $B \perp  ab$ plane. 
A single broad peak is observed below 4.4 T, while a double-horn-type line shape is observed above 4.7 T. 
The transition field from phase I to II is estimated to be 4.5 $\pm$ 0.2 T, 
consistent with the field of 4.5 $\pm$ 0.5 T determined in the polycrystalline sample \cite{MYoshida1}. 
Even in crystal A, the transition shows a small width. Such width is usually attributed to the coexistence region. 
However, the spectra near 4.5 T could not be reproduced simply by a sum of the spectra in phase I and II. 

\begin{figure}[t]
\includegraphics[width=0.7\linewidth]{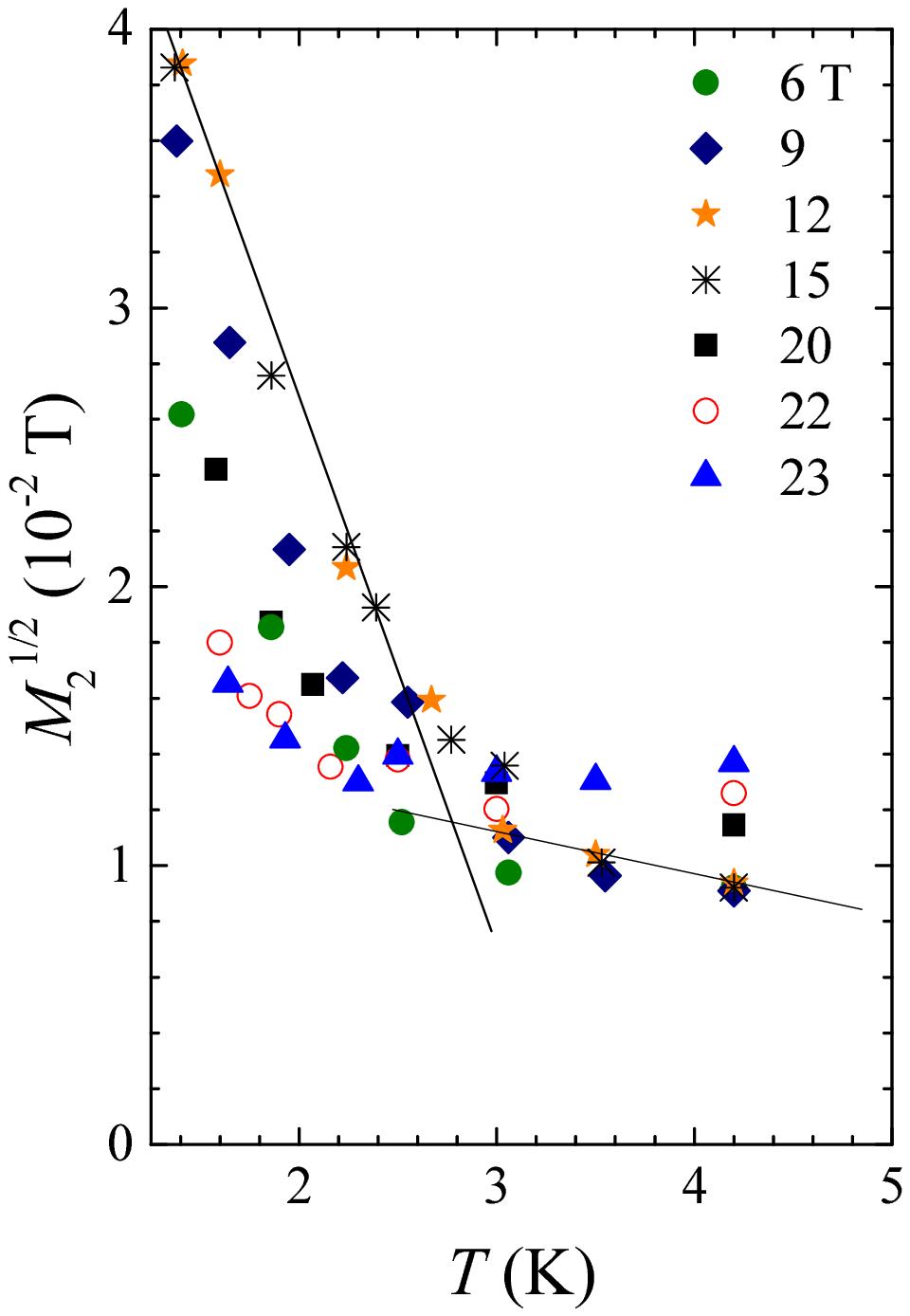}
\caption{Temperature dependence of $(M_2)^{1/2}$ for crystal A at various magnetic fields in $B \perp  ab$ plane.}
\label{width}
\end{figure}

The phase boundaries between phase II and the paramagnetic phase are not clear in the temperature dependence of 1/$T_1$ 
as shown in Fig. 3 of the main text. Therefore, we determined the boundaries from the temperature dependence of 
the line width. Because the quadrupole satellite lines are broadened by powder averaging, the temperature dependence 
of the center line can be used to determine the transition temperature in polycrystalline samples \cite{MYoshida1, MYoshida3}. 
Unfortunately, it is difficult to obtain the temperature dependence of the center line in single crystal measurements, 
because the center and quadrupole satellite lines overlap each other. 
Therefore, we used the temperature dependence of the second moment $M_2$ obtained by 

\begin{eqnarray}
M_2 = \int (B_{\mathrm{int}}-M_1)^2I(B_{\mathrm{int}})dB_{\mathrm{int}}, 
\end{eqnarray}
where $M_1 = \int B_{\mathrm{int}}I(B_{\mathrm{int}})dB_{\mathrm{int}}$ is the center of gravity 
and $I(B_{\mathrm{int}})$ is the NMR spectrum normalized as $\int I(B_{\mathrm{int}})dB_{\mathrm{int}}$ = 1. 
Figure~\ref{width} shows the temperature dependence of $(M_2)^{1/2}$ for crystal A at various magnetic fields 
in $B \perp  ab$ plane. We can see that $(M_2)^{1/2}$ increase with decreasing temperature, although the transition 
is not sharp compared with the results of the center line for the polycrystalline sample \cite{MYoshida1,MYoshida3}. 
We determined the transition temperature by the cross point of the two linear lines fit to the data in the paramagnetic 
and ordered phases, respectively. For example, the two lines for the data at 12 T are shown in Fig.~\ref{width}. 

\subsection{Sample dependence of the spectra in phase II}

\begin{figure}[t]
\includegraphics[width=0.7\linewidth]{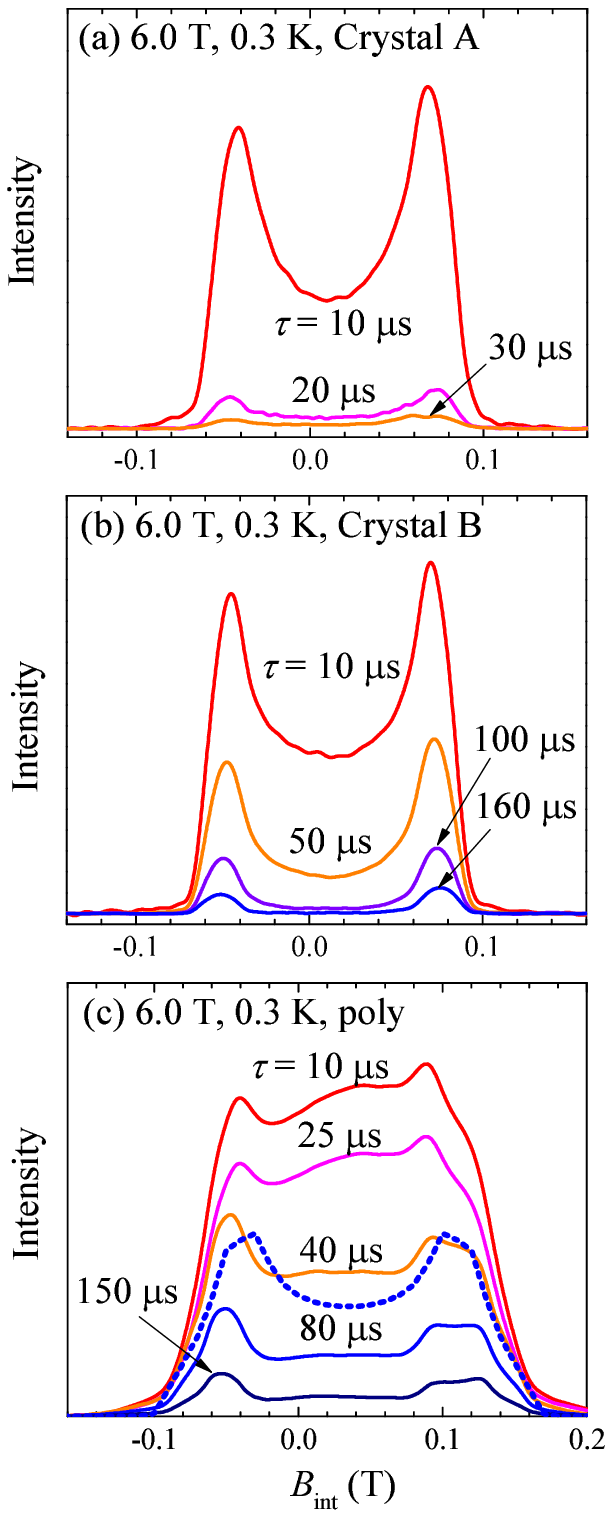}
\caption{$\tau$ dependences of the NMR spectra at 0.3 K and 6.0 T for (a) crystal A and 
(b) crystal B in $B \perp  ab$ plane, 
and (c) the polycrystalline sample taken from Ref. \cite{MYoshida2}.}
\label{spII}
\end{figure}

The NMR results in phase II slightly depend on the sample quality. Figure~\ref{spII} shows the $\tau $ dependences 
of the NMR spectra for the three samples. As shown in Fig.~\ref{spII}(a), 1/$T_2$ is very large in phase II for crystal A, 
indicating that slow spin fluctuations which strongly enhance 1/$T_2$ \cite{MYoshida1} also exist in the SDW state. 
As shown in Fig.~\ref{spII}(b), the same spectral shape is observed in crystal B, while 1/$T_2$ becomes much smaller. 
Although the origin of the fluctuations is unclear at present, imperfection of crystals seems to suppress the fluctuations 
in the SDW sate. 

Figure~\ref{spII}(c) shows the $\tau$ dependence of the NMR spectra for the polycrystalline sample \cite{MYoshida2}. 
The dotted line is a powder pattern spectrum calculated by using the anisotropy of the hyperfine coupling in the paramagnetic phase, 
assuming that the width of the double-horn spectrum depends on this anisotropy in the paramagnetic phase. 
This assumption is valid for the collinear SDW order, where the ordered moments are always parallel or antiparallel to 
the external field. The dotted line reproduces the characteristic features of the spectra after $\tau$ = 40 $\mu$s. 
In the polycrystalline sample, the fast and slow decay components are observed and the former 
becomes negligible after $\tau$ = 40 $\mu$s \cite{MYoshida2}. Therefore, the slow decay component corresponds to the SDW order. 
The spin-echo decay rate for the slow decay component in the polycrystalline sample is 
close to that for crystal B. This fact is consistent with the expectation that crystal B is medium quality. 
In the polycrystalline sample, the fast decay component that vanishes until $\tau$ = 40 $\mu$s has a Gaussian-like line shape \cite{MYoshida2}. 
This component seems to be absent in the single crystals used in the NMR measurements, 
which show only the SDW component as shown in Figs.~\ref{spII}(a) and~\ref{spII}(b). 
The Gaussian-like line shape indicates that the fast decay component originates from imperfection of the polycrystalline sample 
and corresponds to the disorder component defined in section A. 
We further discuss the role of the disorder component in section F. 

The calculated SDW spectrum in Fig.~\ref{spII}(c) is different from the SDW spectrum discussed in Ref. \cite{MYoshida3}. 
In the former case the ordered moments are always parallel or antiparallel to the external field, 
while in the latter case the ordered moments are fixed to a crystalline axis. 

\subsection{Stretch exponent $\beta$}

\begin{figure}[t]
\includegraphics[width=0.8\linewidth]{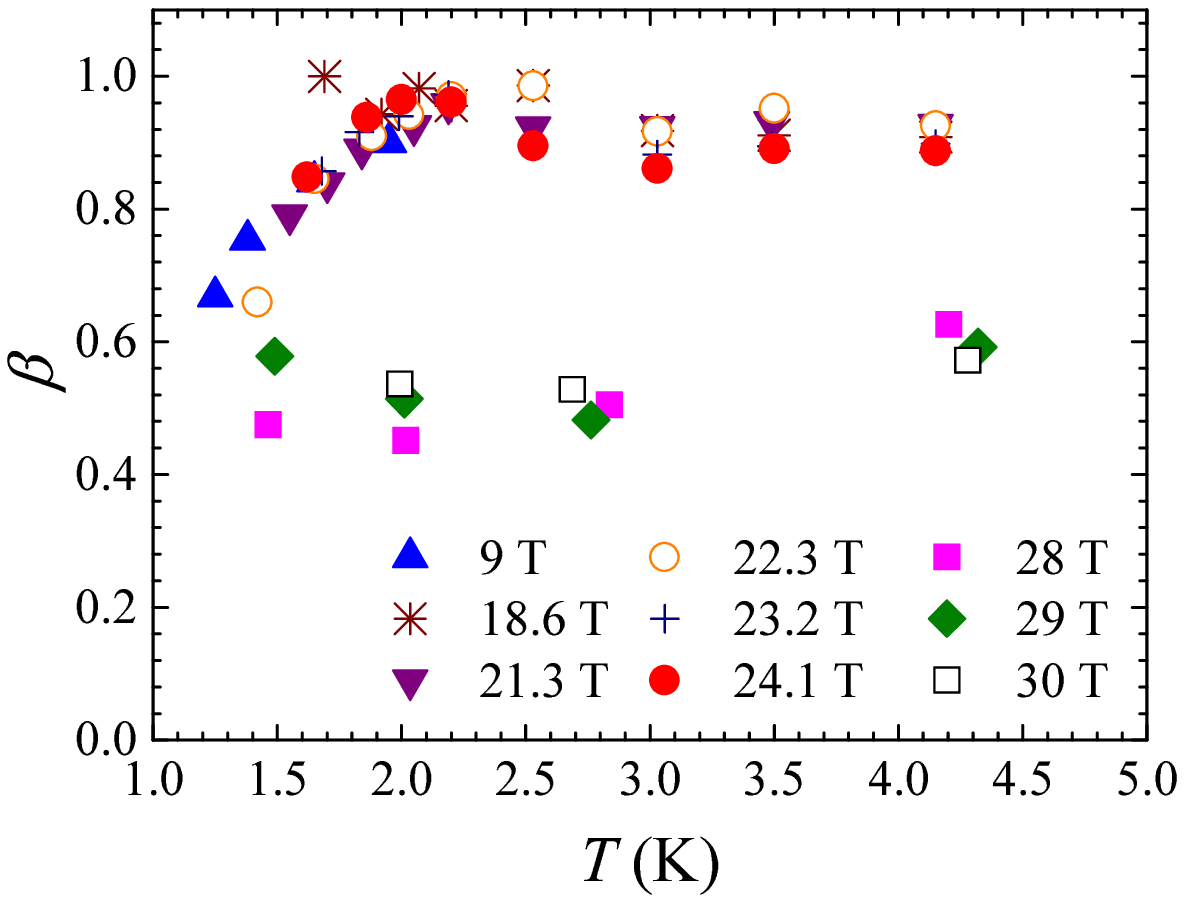}
\caption{Temperature dependence of the stretch exponent $\beta$ for the 1/$T_1$ data shown in Fig. 3 of the main text.}
\label{beta}
\end{figure}

Figure~\ref{beta} shows the temperature dependence of the stretch exponent $\beta$ for the 1/$T_1$ data shown in Fig. 3 of the main text. 
At 9 T, the recovery curve can be fit to the single exponential function above 2 K, while 1/$T_1$ shows a distribution below 2 K. 
The temperature dependences of $\beta$ at 18.6-24.1 T are quite similar to that at 9 T, indicating an appearance of internal fields below 2 K. 
These are consistent with the temperature dependences of the line widths shown in Fig.~\ref{width}. 
Above 28 T, 1/$T_1$ shows a distribution even at 4.2 K in the paramagnetic state. 
In general, effect of crystalline defects on 1/$T_1$ becomes significant in a gaped state. 
Then, the strong external field might induce inhomogeneous $B_{\mathrm{int}}$ near defects, 
which could provide inhomogeneity of 1/$T_1$. 

\subsection{Two components in 1/$T_1$ at the plateau region in crystal A}

\begin{figure}[t]
\includegraphics[width=0.8\linewidth]{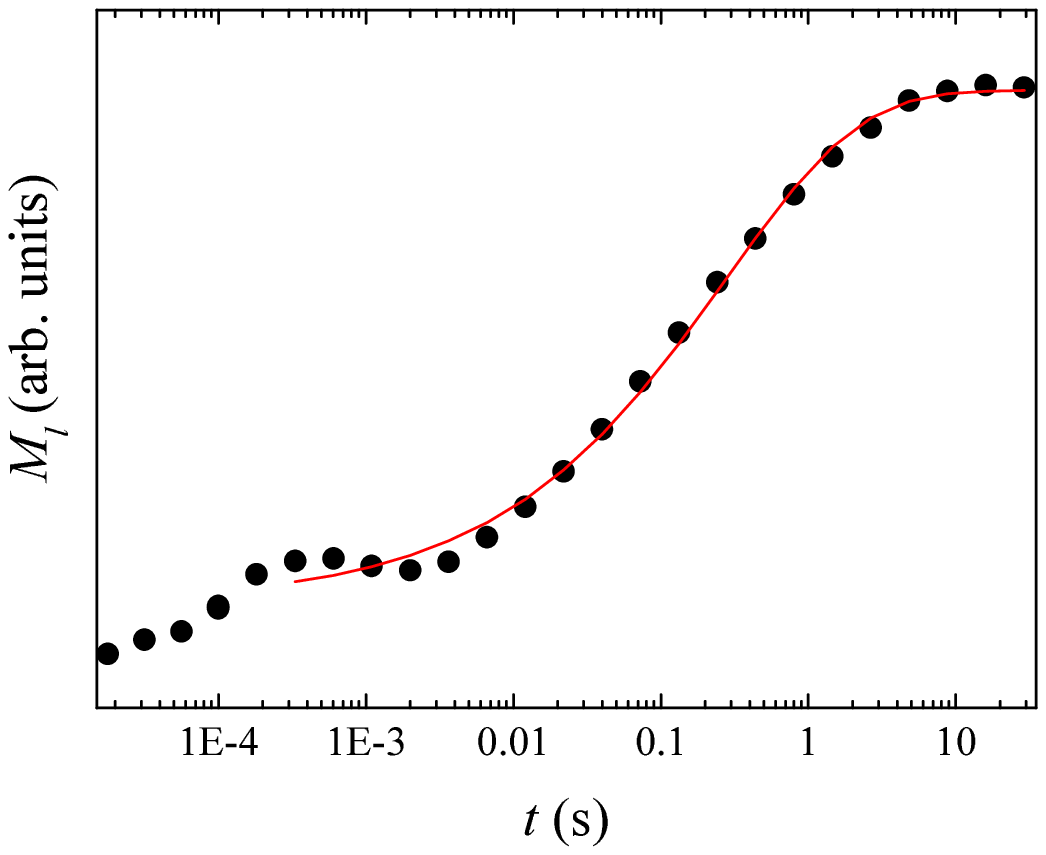}
\caption{Recovery curve (points) and the fit (line) for the $T_1$ measurement in crystal A at $\tau$ = 10 $\mu$s and 1.5 K at 28 T in $B \perp  ab$ plane.}
\label{recovery}
\end{figure}

The two components in crystal A can be clearly seen in the spin-echo decay curve at 28 T 
below 1.5 K as shown in Fig 4(a) in the main text. The two components can also be seen 
in the recovery curve for the 1/$T_1$ measurements above 28 T. Figure~\ref{recovery} shows the recovery curve 
for crystal A at $\tau $ = 10 $\mu$s and 1.5 K at 28 T in $B \perp  ab$ plane. 
We can see the two components; the intensity of one component is recovered within 10$^{-3}$ s, while the intensity 
of the other component is recovered near 10 s. The values of 1/$T_1$ in Fig. 3 of the main text are obtained 
by fitting the slow decay component to the stretch exponential function as shown by the solid line in Fig.~\ref{recovery}. 
It is noted that in general spin diffusion effects might provide a fast relaxation component. 
However, the observed two components should correspond to the two distinct environments of internal fields, 
because they are affected by different $B_{\mathrm{int}}$ as shown in Fig. 4(b) of the main text. 

\subsection{Possible role of the two layers in the polycrystalline sample}

\begin{figure}[b]
\includegraphics[width=0.8\linewidth]{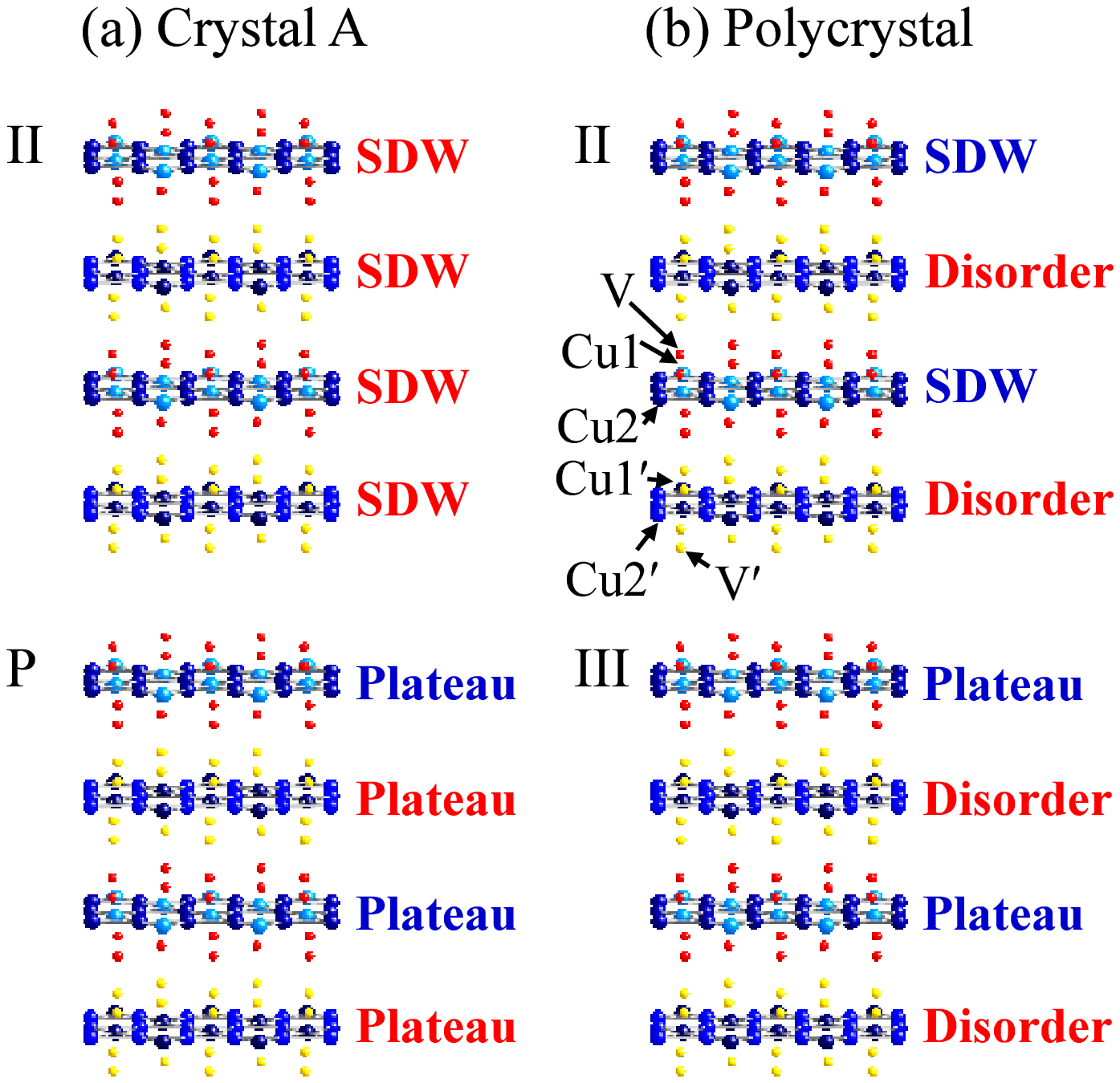}
\caption{Schematic structure of (a) phase II and P state for crystal A and (b) phase II and III 
for the polycrystalline sample. SDW, Plateau, and Disorder indicate the spin states with large (red) or small (blue) 1/$T_2$.}
\label{model}
\end{figure}

In phase II for the polycrystalline sample, $B_{\mathrm{int}}$ of the disorder component 
has a Gaussian-like distribution \cite{MYoshida2}, while the other component is consistent 
with the SDW order as discussed in section C. The disorder component persisted even in phase III 
for the polycrystalline sample, resulting in the sample dependence of the magnetization curves as discussed in section A. 
A previous study argued that a heterogeneous spin state consisting of two spatially alternating Cu spin systems 
in the kagome layer is realized \cite{MYoshida2}. There are two experimental results which support a heterogeneous 
spin state in the polycrystalline sample. One is the same number of the V sites for the two components, 
which indicates a microscopic superstructure. The other is the similar temperature dependences of 1/$T_2$ for the two components, 
which excludes a possibility of a macroscopic phase separation. In the previous study \cite{MYoshida2}, 
a magnetic superstructure within the kagome layer was proposed, because there is a unique kagome layer 
in the $C2/m$ structure. The interplane interaction is expected to be weak, which is unlikely to cause a symmetry breaking 
on the equivalent layers. However, the single crystal X-ray studies revealed the $P2_1/a$ structure below 155 K, 
in which there are two inequivalent kagome layers. The crystallographic difference between the two layers is likely to 
stabilize different spin states in the two layers at low temperatures. 

Let us further examine the two layer scenario by using Figs.~\ref{model} (a) and (b). 
In phase II for crystal A, only one component with the large 1/$T_2$ is observed as shown in Fig.~\ref{spII}(a). 
This result indicates that a uniform SDW order is realized in phase II for crystal A as shown in Fig~\ref{model}(a), 
in spite of the crystallographic difference between the two layers. In P state for crystal A, 
the two components with the different 1/$T_2$ are observed as shown in Fig. 4 in the main text. 
As discussed in the main text, the large difference of 1/$T_2$ can be attributed to the difference of the critical fields. 

A likely origin of imperfection in crystal B and the polycrystalline sample is disorder of the crystal water molecules 
between the kagome layers or imperfection of the structural transitions. In phase II for crystal B, 1/$T_2$ is much smaller 
compared with that for crystal A as shown in Fig.~\ref{spII}(b). This result may be attributed to a disorder effect on 
the interlayer coherence. In phase II for the polycrystalline sample, such an effect might stabilize different spin states as 
shown in the upper panel of Fig.~\ref{model}(b). Assuming that only the SDW component reaches the 1/3 plateau in phase III 
as shown in the lower panel of Fig.~\ref{model}(b), the magnetization step at 25 T can be explained. 
The step at 46 T might be attributed to the increase of $M$ of the disordered layer toward the 1/3 plateau, 
although $M$ for the polycrystalline sample exceeds the 1/3 of the saturated magnetization above 52 T.

\end{document}